\documentclass[journal,twocolumn]{IEEEtran}
\ifCLASSINFOpdf
  \usepackage[pdftex]{graphicx}
  \graphicspath{{graphics/}}
\else
\fi

\usepackage{mhchem}
\usepackage[normalem]{ulem}
\usepackage{textcomp}

\usepackage{array}
\newcolumntype{P}[1]{>{\centering\arraybackslash}p{#1}}
\newcolumntype{M}[1]{>{\centering\arraybackslash}m{#1}}
\usepackage{floatrow}
\usepackage{threeparttablex}
\usepackage{cancel}

\usepackage{adjustbox,longtable,multirow}
\usepackage{subfig}

\usepackage{booktabs}
\begin{document}
%
\title{Evolving Intelligent Reflector Surface towards 6G for Public Health: Application in Airborne Virus Detection}
%
%
%

\author{
Harun \v{S}iljak, 
Nouman Ashraf,
Michael Taynnan Barros, 
Daniel Perez Martins,
Bernard Butler,
Arman Farhang,
Nicola Marchetti,
Sasitharan Balasubramaniam
\thanks{H\v{S} and NM are with CONNECT Centre, Trinity College Dublin, Ireland; NA, DPM, BB and SB are with the Telecommunication Software \& Systems Group, Waterford Institute of Technology, Ireland; MTB is with the University of Essex, United Kingdom; AF is with Maynooth University, Ireland.
This publication has emanated from research conducted with the financial support of Science Foundation Ireland (SFI) and the Department of Agriculture, Food and Marine on behalf of the Government of Ireland under Grant Number [16/RC/3835] - VistaMilk, and is co-funded under the European Regional Development Fund for the CONNECT research centre (13/RC/2077).
Corresponding author (H\v{S}): harun.siljak@tcd.ie}
}
\maketitle

\begin{abstract}

 While metasurface based intelligent reflecting surfaces (IRS) are an important emerging technology for future generations of wireless connectivity in its own right, the plans for the mass deployment of these surfaces motivate the question of their integration with other new and emerging technologies that would require mass proliferation. This question of integration and the vision of future communication systems as an invaluable component for public health motivated our new concept of Intelligent Reflector-Viral Detectors (IR-VD). In this novel scheme, we propose deployment of intelligent reflectors with strips of receptor-based viral detectors placed between the reflective surface tiles. Our proposed approach encodes information of the virus by flicking the angle of the reflected beams, using time variations between the beam deviations to represent the messages. This information includes the presence of the virus, its location and load size. The paper presents simulation to demonstrate the encoding process based on varying quantity of virus that have  bound onto the IR-VD.
\end{abstract}

\begin{IEEEkeywords}
6G communications, Virus detectors, Airborne virus,  Intelligent Reflector Surfaces, Molecular communications, COVID-19.
\end{IEEEkeywords}

\IEEEpeerreviewmaketitle

\section{Introduction}

\IEEEPARstart{T}{he} 
recent COVID-19 pandemic has put the world on its heels with numerous countries around the world facing lockdowns that have affected the global economy. The virus itself posed significant challenging questions to the scientific community, and this includes, (i) how does the virus affect the human body, (ii) how does it spread in indoor and outdoor spaces and (iii) how to provide rapid and efficient virus sensing mechanisms.

Since this is a global challenge, a collaborative effort is required from scientists that goes beyond virologists, immunologists and public health experts. Researchers in information and communications technology (ICT) have recently begun working towards solutions that can help curb the spread of the virus, through contact tracing applications on mobile devices. While 6G networks are expected to chiefly focus on high capacity and latency \& reliability requirements for applications as diverse as pervasive connectivity and Industry 4.0, here we join the growing community \cite{ziegler20206g} of voices calling for its paramount importance in enabling novel and efficient approaches to public health. In this paper, we propose a new solution where wireless communication systems can play a role in detecting the virus shed by infected individuals. 
This is based on the use of Intelligent Reflective Surfaces (IRS) that have been proposed for reflecting, refracting and diffracting electromagnetic waves in the mm-wave and terahertz spectrum. Such electromagnetic waves suffer from numerous  effects that lead to unreliable communication between the transmitter and receiver. Firstly, the challenge of high spreading loss means that high-gain, highly directive antennas are needed to extend range. Consequently, the signals appear as “pencil-thin” beams that are highly directive, requiring Line-of-Sight (LoS) visibility (hence no obstacles) between the communication points. This is closely tied with the effect of atmospheric conditions (e.g., raindrops and fog (water vapour)) that absorb the incident energy at those frequencies and so can significantly attenuate the signals, thereby further reducing their range. IRS \cite{Liaskos2018anewwireless, ashraf2020extremum} have been proposed to counter these challenges, by redirecting and reflecting beams to the mobile devices, to ensure non-LoS signal coverage. The introduction of IRS will require infrastructure changes, where this could come in the form of wallpapers that integrate electronics and nanotechnology to ensure compactness, flexibility and scalability in the installed reflectors. 
With this pervasive infrastructural intervention, it is reasonable to plan other generally useful features that could be symbiotically tied to IRS. An IRS can co-exist with very dense sensors which uses the reflector for coding spatiotemporal information about measurements, updating information in real time and, if needed, integrate with the smart building functionalities.

\begin{figure*}[ht!]
    \begin{center}
        \includegraphics[width=0.9\columnwidth]{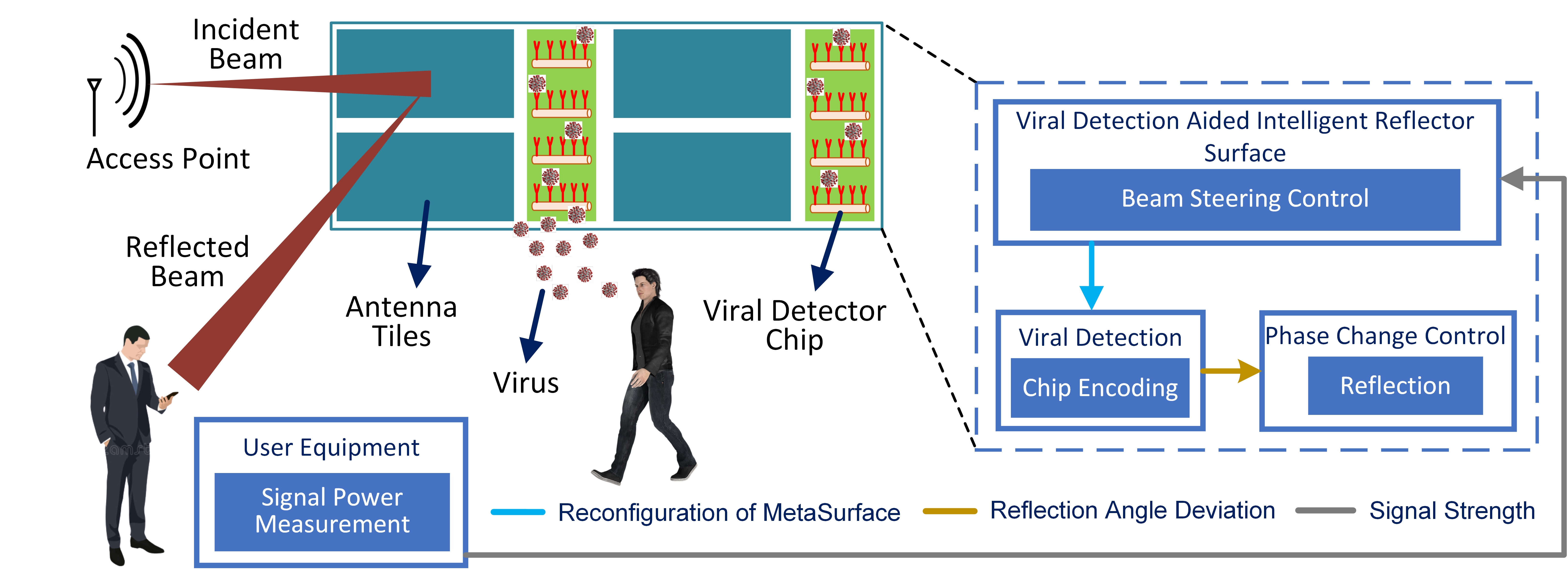}
    \end{center}
    \caption{Setup of Intelligent Reflector - Viral Detectors (IR-VD) with an array of patch antennas integrated with adhesive viral detector strips, as well as protocol modules for controlling the beam forming and manipulation for information encoding.}
    \label{fig:Overall}
\end{figure*}

Our proposed solution is a symbiotic integration of viral detectors with the IRS. In such an Intelligent Reflector - Viral Detectors (IR-VD), adhesive strips with chemical receptors capture virus particles that bind to the strips to perform detection. 
Upon sensing virus particles that have bound to the strips, the 
unit cells will reflect signals with an embedded code to provide information about the virus, and comes in the form of phase manipulation that results in beam flicking encoding a binary string, which is detected by the receiver by sensing the power deviation of the signals. Patches that are not adjacent to such virus-affected areas will not have this embedded code. Once detected by the mobile device, the embedded code and its location will indicate the zone of the wall that is contaminated with the virus. The flexibility of the adhesive strips applicable to most virus (e.g., SARS-CoV-2) that can have engineered receptors. This in turn will result in a different code that is embedded into the signals reflected from the IR-VD. 

In this paper, we present the design of IR-VD, motivation for its development, and simulation of its performance which both verifies the principle of operation and explains it practically. We give a contextual overview of viral detection, as well as the convergence of different technologies and means of communications that have enabled development of the concept, and discuss potential challenges and future work.


\section{Airborne Virus Background}\label{sec:airborne_virus}
Indoor environments can expose humans to a variety of airborne viruses, some of which may cause respiratory diseases. These pathogens can be transmitted through air in the form of liquid droplets or airborne particles \cite{hurst2007manual}. Due to their small sizes ($1-40\,\mu$m when coughing) and low settling velocity (up to $3.1\times10^{-3}$ m/s for $10\,\mu$m particles), these viruses can remain in the air, increasing their likelihood of infecting other people within an indoor environment \cite{hurst2007manual,fronczek2015biosensors}. The virus survivability in the air is also related to environmental factors, such as temperature, relative humidity (RH) and surrounding biological materials that can absorb the virus \cite{ijaz2016generic}. From these factors, higher temperatures decrease the virus survivability while the surrounding biological material increases it. In this case, the higher survivability is related to the physical protection against environmental changes provided by the biological material (i.e., mucus) that encapsulates the viruses \cite{ijaz2016generic}. On the other hand, RH values affect the survivability of each virus family differently. For example, most respiratory viruses, including influenza, survive longer in environments with RH between $20\%$ and $30\%$. However, non-enveloped viruses (e.g., polio virus) show higher survivability at higher RH that is between $70\%$ and $90\%$ \cite{ijaz2016generic}.

\begin{figure*}[ht!]
    \begin{center}
        \includegraphics[width=\textwidth]{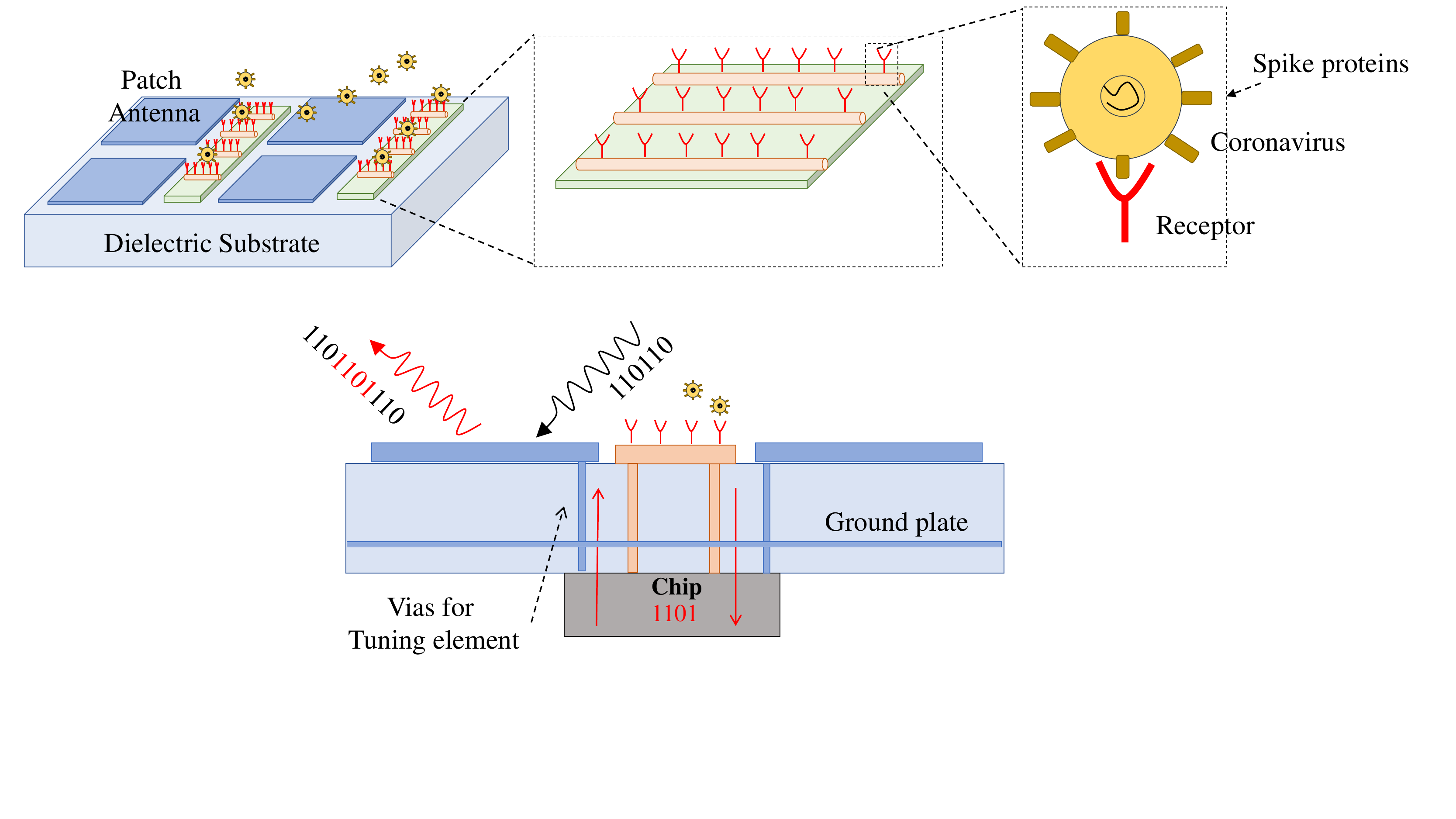}
    \end{center}
   \vspace{-2 cm}
    \caption{Overview of the IR-VD, with an expanded view of the viral detector strip, showing the conductive tubes and the viral receptors. The viral receptor is applied on the surface of the conductive tubes to bind and detect the virus. The lateral projection of the intelligent reflector illustrates the generation and embedding of a binary code (highlighted in red) on the reflected electromagnetic wave whenever a virus is detected.}
    \label{fig:Devices}
\end{figure*}

Viral and droplet/particle sizes are also related to the different infected areas of the respiratory tract. For example, when inhaled, larger particles (greater than $5\,\mu$m) deposit on the upper respiratory tract while smaller particles deposit in the lower part of the lungs and can cause severe pneumonia \cite{hurst2007manual}. The infected people will then become the new source of viral aerosols and can spread them through coughing, sneezing or speaking when are close to other people. In addition to these, there are other sources of viral aerosols in the home environments, such as flushing toilets (liquid droplets coming from the toilet can carry viral particles excreted from the human body) and changing diapers (by manipulating diapers some viral particles can be thrown in the air), and in laboratory settings, e.g. shaking, homogenisation and sonication of materials that contain viral particles \cite{hurst2007manual}. The campaign around social distancing can reduce viral aerosols and reduce the propagation of virus to infect other people \cite{hurst2007manual}  \cite{fronczek2015biosensors}. Therefore, the air quality has become a public health and safety concern due to the annual numbers of respiratory infections, specially in the recent years with the SARS and COVID-19 pandemics, which has driven investigation into new methods to improve airborne virus detection and identification  \cite{ijaz2016generic}. In the next section, we introduce a new solution to monitor the presence of airborne viruses within an indoor environment, which can be applied for commercial or domestic purposes to help detect airborne as well as viral droplets.

\section {Intelligent Reflectors for Viral Detection}

As described in the introduction, the future wireless communication systems envisions a new form of  IRS infrastructure  placed on the wall that are used to reflect high frequency spectrum signals to devices. Our proposed approach is to make use of this new infrastructure  to detect viral particles, enabling it to have dual function: high-speed wireless communication and sensing airborne viral and communicating to mobile devices.  
This makes viral sensing and IRS a good match, since virus sensing can be integrated with other prominent wireless communication systems of the future, which is appealing from a deployment perspective as well as a future sensing technology: pervasive, reliable, high-tech entities that not only coexist, but support one another. In the context of proliferation incentives, they also form a symbiotic relationship: the push for better connectivity would bring health protection, and vice versa, contributing to wide market penetration.

\subsection{Viral Detection Beam Steering Protocol}

\begin{figure}[htb]
\centering
\includegraphics[width=.93\columnwidth, height=360 pt]{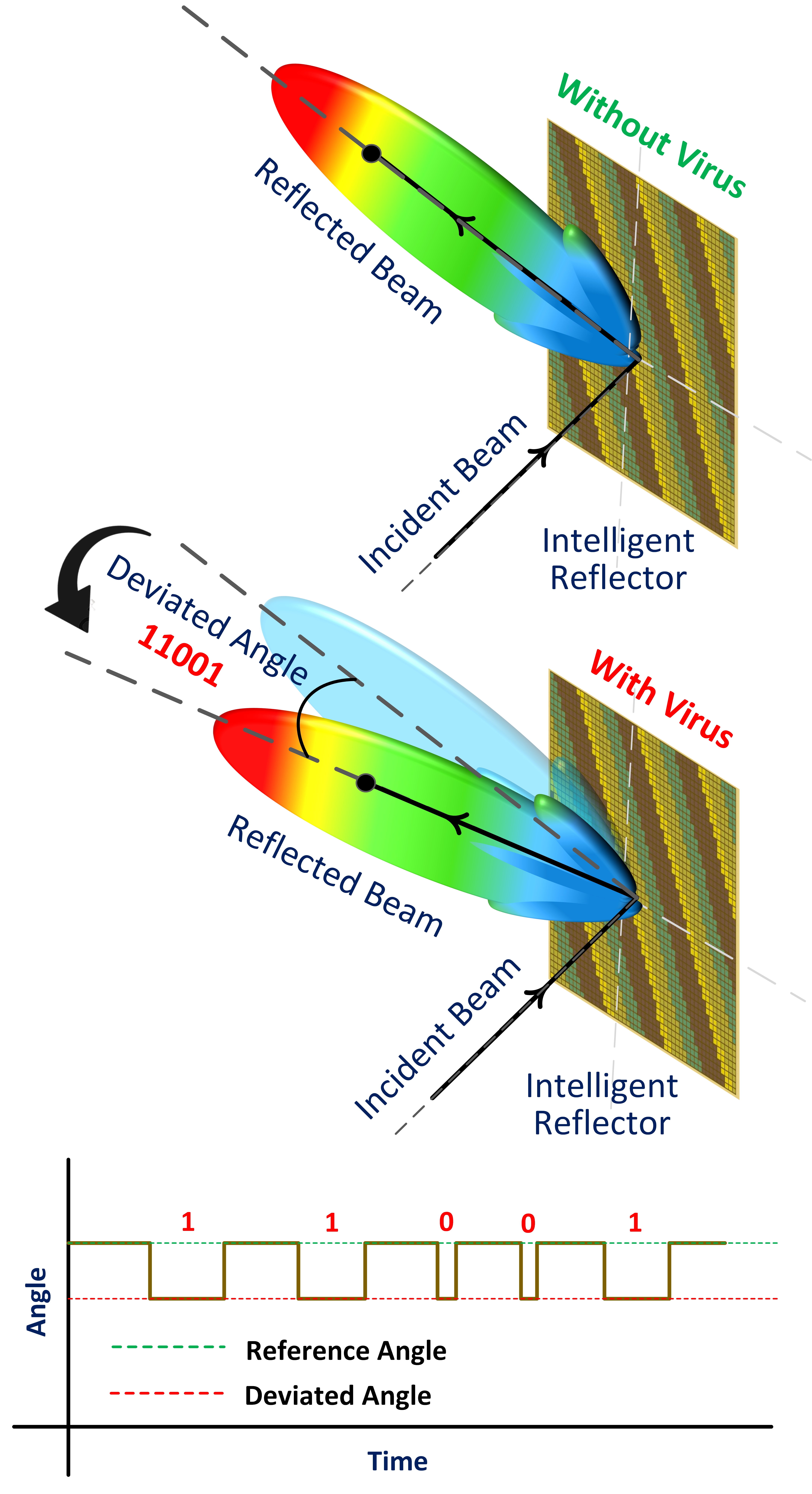}
\caption{The proposed approach for information encoding via phase change of the reflected beam. In absence of virus, the beam reflects towards the receiver, while presence of virus encodes the information by flicking the beam. Time variations between flicking of the beam generates the information bits about the presence, location and concentration of the virus.}
\label{fig:blockdia}
\end{figure}

IRS functionalities can be further extended to detect airborne virus within an indoor environments based on people coughing or sneezing by adding viral detector strips, illustrated in Fig. \ref{fig:Devices}. The strips, combined with the components found in most IRS which includes patch electromagnetic antennas, a dielectric substrate, a ground metallic material, and a chip that controls the phase changes of the reflected signals constitute the IR-VD. The adhesive viral detector strips will be placed in the spaces between the patch antenna tiles and contain functionalized receptors that correspond to the proteins on the virus membrane. These receptors can detect single or multiple families of viruses. For example, human $\alpha$-2,6-linked sialo-saccharides have been reported to bind influenza virus family, making these sugars suitable candidates for the receptors \cite{de2014role}. In terms of integrating with the conductance change that interfaces to the chip, we can foresee the usage of electrostatic-based detection using protein-ligand response that changes the conductance in a substrate with response time around 2 minutes for viruses such as Influenza, A/HN1, H5N1 AND SARS  \cite{fronczek2015biosensors}. 
This modular design will enable the providers to customise the reflectors to their needs, such as focusing on the detection of the most dangerous families of viruses, in case of a viral outbreak or pandemic. The process of viral detection and monitoring can be described as follows: upon viral contact with the adhesive receptor strips, the conductance along the tubes containing the receptors will change, indicating its presence on a particular section of the IR-VD. 
The change in conductance will be detected by the chip that controls the beam steering, and will encode information by flicking the phase change of the signals, as depicted in Fig. \ref{fig:blockdia}. The encoded information will depend on the duration between the phase changes (e.g., a digital 1 bit will result in a longer durations between the steered beams compared to digital 0 bit)  \cite{ashraf2020extremum}. Therefore, the flicking of the beams will need to be detected by the receiver and will also need to detect the timing between the power changes as the beams are flicked. The encoding of the information can include the type of virus, as well as the density on the wall, and possibly the location on the IR-VD. 

\begin{table*}
\begin{ThreePartTable}
\begin{tabular}{M{0.1\textwidth}M{0.1\textwidth}M{0.18\textwidth}M{0.1\textwidth}M{0.08\textwidth}M{0.12\textwidth}M{0.1\textwidth}}
\caption{Literature review on IRS and quantity of viral detectors per tile of antenna patches that can be placed in these solutions.}
\label{tb:metasurfaces}\\
    \hline
    {\bf Metasurface Solution} & {\bf Frequency of Operation} & {\bf Free Spacing} & {\bf Tile Size} & {\bf Materials Used} & {\bf Steering Granularity} & {\bf Viral Detectors per Tile ($2\times2\,\text{mm}^2$)}
    \\ \hline
    \cite{dai2020reconfigurable} & $2.3$ GHz and $28.5$ GHz& $13\times(37,37,50,50)\,\text{mm}^2$& $50 \times 50\,\text{mm}^2$ & FR4 substrate & $-60^{\circ}+60^{\circ}$ & $\approx565$ \\
    
    \cite{li2019intelligent} & $2.4-2.5$ GHz &  $17\times(37,36,17.4,26.6)$ $\text{mm}^2$  & $54\times54\,\text{mm}^2$ & F4B/FR4 substrate & N/A & $\approx726$ \\
    
    \cite{naqvi2019beam} & $2.6$ GHz & $16\times6\,\text{mm}^2$ (two areas) & $20 \times 20\,\text{mm}^2$ &  Rogers RO3003 / PLA / Water & $-20^{\circ}-+20^{\circ}$ & $48$
    
    
    \\
   
    \hline
\end{tabular}
\end{ThreePartTable}
\end{table*}

\subsection{Viral Detectors Strips for IRS}

Since the introduction of IRS, a number of different solutions and configurations have been proposed. In Table \ref{tb:metasurfaces} we describe three IRS solutions based on their frequency of operation, tile size material used and steering granularity. Moreover, we describe the free spacing on the tile (i.e., space that is not occupied by the patch antenna or other electronic device), and the density of viral detectors that can be placed on the tile depending on this free spacing. In \cite{dai2020reconfigurable}, a novel antenna structure is proposed to individually control the phase of the reflected signal by configuring different RF paths. Each tile is composed of three reflective paths arranged in a T-shaped configuration with a fourth dummy path just to maintain a symmetric structure. For this solution, the antenna patch is placed in the center of the IRS tile, and the following spaces are free and can be used to placed the viral detectors: $13\times50\,\text{mm}^2$ for the left and right sides, and $13\times37\,\text{mm}^2$ for the upper and lower sides. Based on these areas dimensions, approximately 565 viral detectors can be placed in a single tile.

A similar structure was proposed in \cite{li2019intelligent}, where an IRS was designed to recognize human gesture and respiration in real-time by transmitting and receiving electromagnetic waves and processing them using an artificial neural network. This solution has plenty of open areas around the antenna patch where the viral detectors can be placed at: $17\times54\,\text{mm}^2$ on the left side, $17\times37\,\text{mm}^2$ the upper side, $17\times36\,\text{mm}^2$ on the lower side (due to the presence of an inductor on the IRS), $17\times17.4\,\text{mm}^2$ and $17\times26.6\,\text{mm}^2$ on the right side (above and below the resistor present on the IRS, respectively). Given the spacing on each tile, this model is able to fill a higher quantity of viral detectors, reaching up to 726 per tile.

Other designs can be considered as well, however they might not posses enough free spacing to place a great number of viral detectors as the squared shape with the centered patched antenna described in \cite{dai2020reconfigurable,li2019intelligent}. For example, we also reviewed the design proposed by \cite{naqvi2019beam}. where small tubes filled with water are used to reflect electromagnetic waves. This IRS achieves different beam phases through a combination of empty and water-filled tubes for the reflectors. This design have small areas between the tubes to allow us to place the adhesive viral detector strips. In this case, the free spacing are two small areas of $16\times6\,\text{mm}^2$, which enables to place $48$ viral detectors.

\section{Simulations}

\begin{figure*}[htb]
\centering

  
    \includegraphics[width=.48\textwidth]{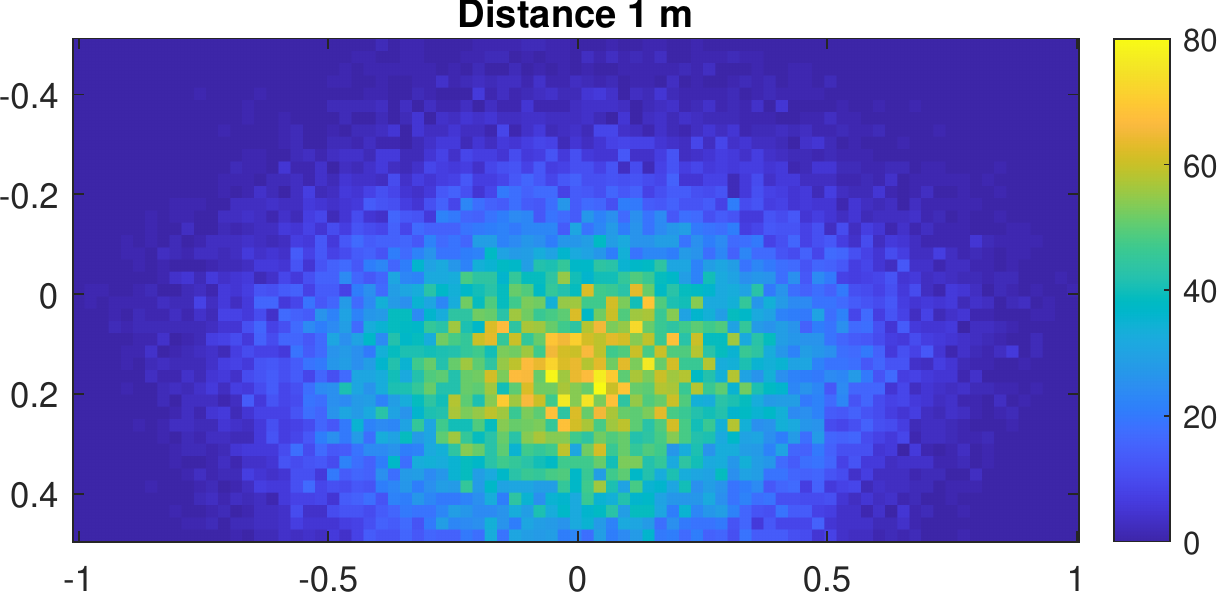}
    \includegraphics[width=.48\textwidth]{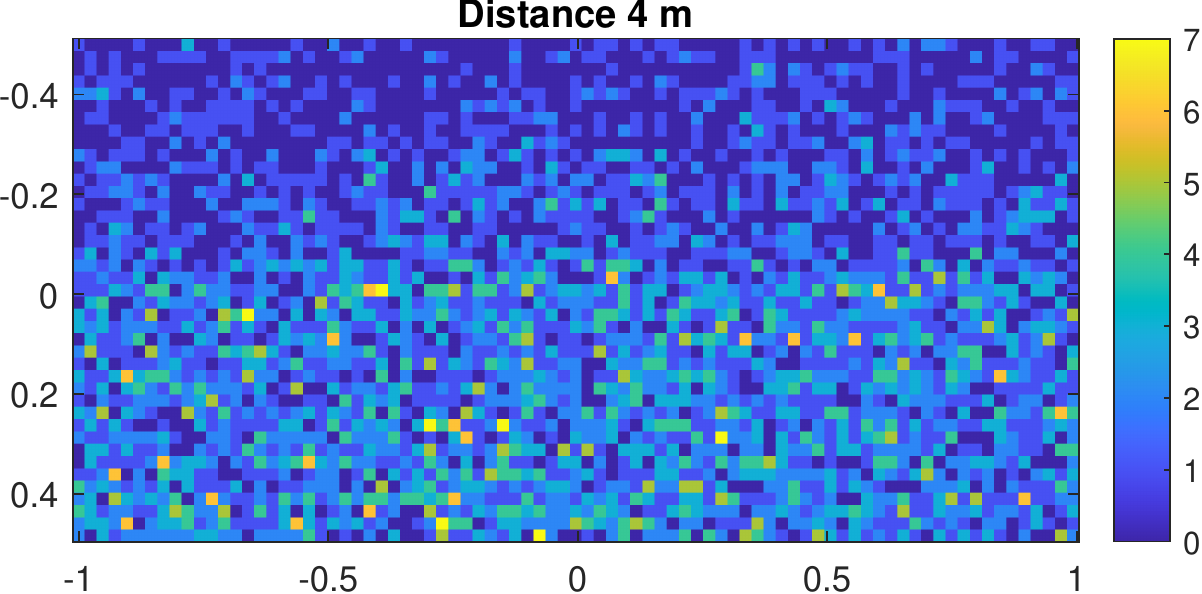}\\
    \subfloat{%
    \includegraphics[width=.49\textwidth, height =138pt]{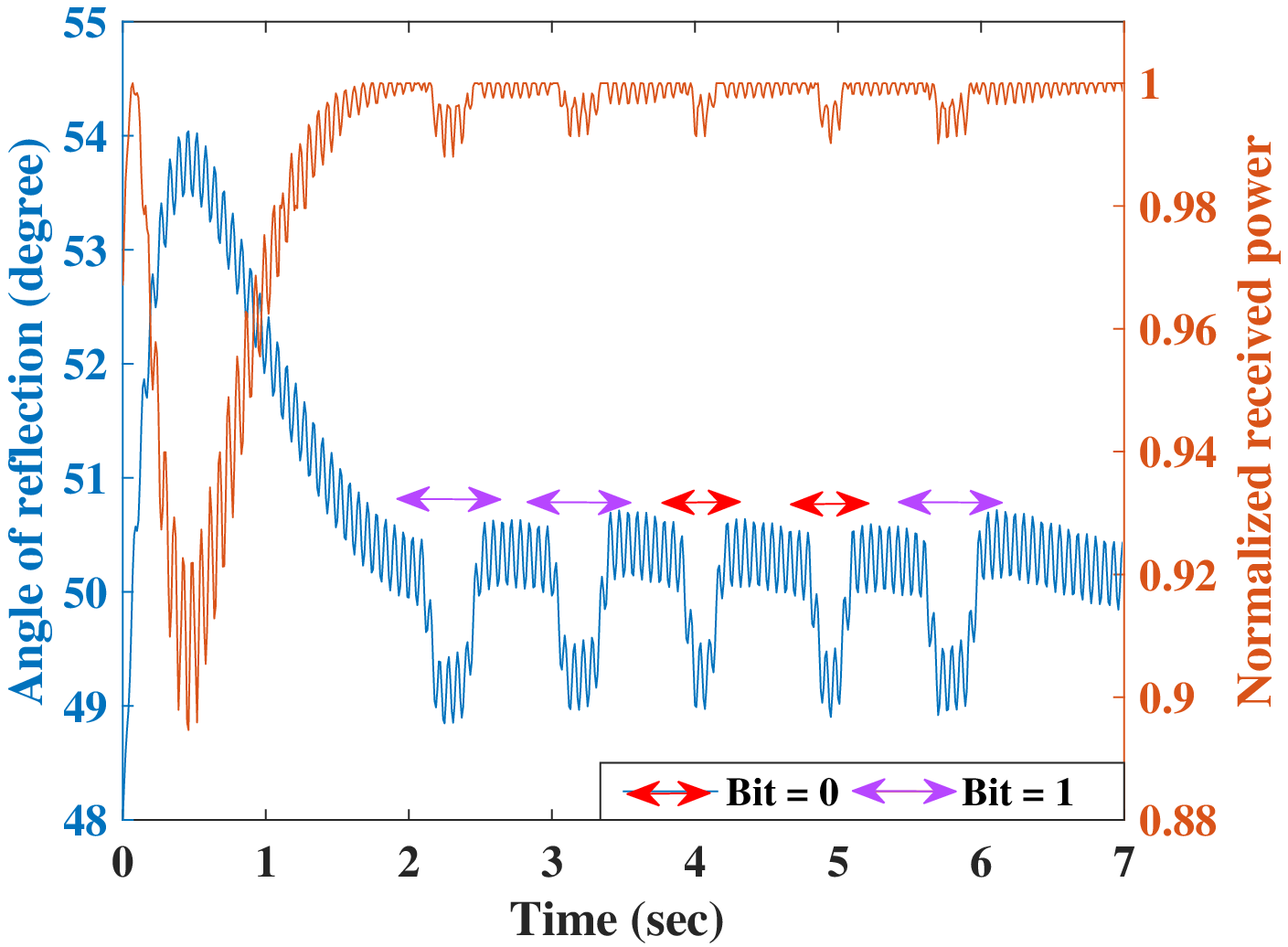}}\hfill
  \subfloat{%
    \includegraphics[width=.49\textwidth, height =138pt]{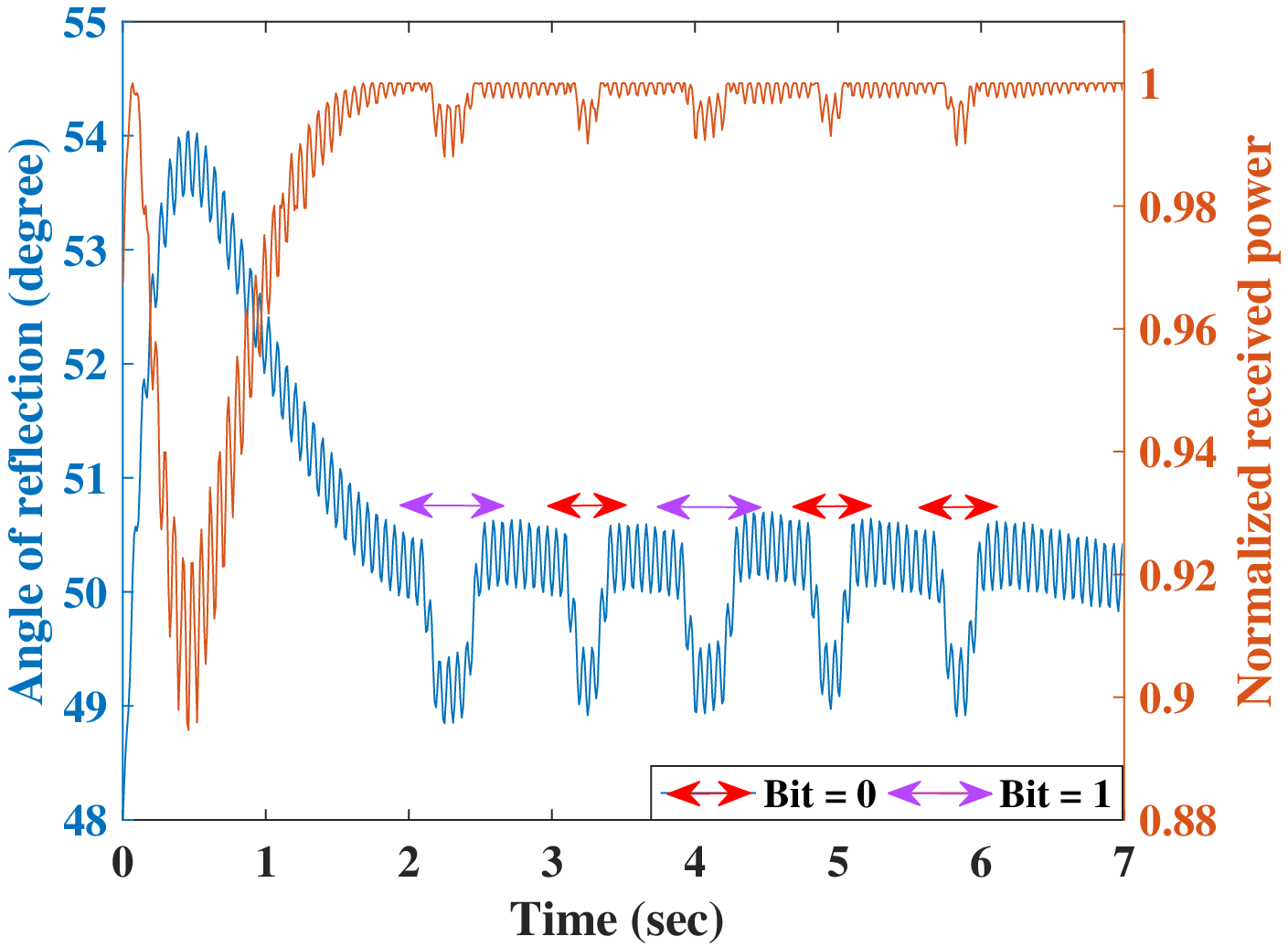}}\hfill
  \caption{Simulation of the IR-VD viral load detection for different concentrations caused by a person sneezing at a distance of 1m or 4m from the IR-VD, interpreted through observed angles of reflection and normalized received signal power}\label{fig:simsneeze}
\end{figure*}



\subsection{Simulation Scenario}

In the scenario shown in Fig. \ref{fig:Overall}, the people inside the room disperse droplets with viral particles (e.g., by sneezing), and the droplets that hit the portion of the wall occupied by our IR-VD is of interest to us. This scenario setting can be interpreted through the concept of an \emph{breath molecular communication model} \cite{khalid2019communication}.

In a breath molecular communication system, we observe the dispersion of aerosol from the respiratory tract of the people present in the space. The droplets with viral content that reaches the IR-VD and binds onto the surface will trigger a message, which is created and encoded into the reflected beam that will flick depending on the data to be transmitted. If observed as a molecular MIMO (multiple input, multiple output) system, where people in the room are the viral sources, and different strips on the reflectors are the receivers, known results about signal detection in such configurations \cite{huang2019rising} show that it is possible to reconstruct which viral sources (i.e. person) transmitted which "message" even if the sources are not spatially distanced. While such an option opens the question of anonymity if identification is done at the level of an individual, it is not hard to devise a use-case in which such information is used through aggregation. For example, if the patches are placed within an air conditioning network serving several rooms at a workplace, knowing which room contains stream that has high probability of containing  virus could help in providing a safe working environment.

\subsection{Viral Molecular Communication and Binding Model}

For the simulated effects of molecular communication sneezing (quick burst transmission in the atmospheric channel towards a dense array of receivers on the wall), the effects of the channel can be modelled in a relatively simple manner. A range of droplets of relatively large mass and volume originating from the same point in space under a range of horizontal and vertical angles and initial velocities travel as projectiles with air drag effect towards the wall. The particular parameters used for this simulation are listed in Table \ref{tb:simpar}. With breathing or speaking, the effects of the dynamic molecular communication channel become more emphasised, with the air movement dictating the motion of small particles with low initial velocities. Once the particles reach the IR-VD, the density of the receptors per tile dictates the sensing capabilities and the resolution of measurement.

\begin{table}[]
\caption{Simulation parameters}
\label{tb:simpar}
\begin{tabular}{ccc}
\hline
\hline
Parameter            & mean                  & st. dev.             \\ \hline
Mass of particles    & $10^{-4}$ g             & $10^{-4}$ g            \\
Initial velocity     & 11.2 m/s              & 3 m/s                \\
Horizontal angle     & $0^{\circ}$           & $18^{\circ}$         \\
Vertical angle       & $-6^{\circ}$          & $12^{\circ}$         \\
Number of particles  & \multicolumn{2}{c}{50,000}                   \\
Receptors per tile   & \multicolumn{2}{c}{160} \\
\hline
Reflector tile & \multicolumn{2}{c}{$24\times24$ cells} \\
Perturbation gain for ESC& \multicolumn{2}{c}{0.5} \\
Perturbation frequency for ESC& \multicolumn{2}{c}{50 rad/s} \\
Initial reflecting angle & \multicolumn{2}{c}{$48^{\circ}$} \\
Angle of receiver w.r.t IR-VD & \multicolumn{2}{c}{$50^{\circ}$}
\\
\hline
\end{tabular}%
\end{table}


The simulated virus binding process is based on the model previously reported in \cite{martins2018computational}. A chemical reaction occurs whenever the virus proteins, i.e. ligands, have physical contact with the host receptors, which in this case they will be the viral detectors, resulting in the binding of these proteins. The strength of this attachment is a design requirement to ensure that the virus will remain bound to the viral detector, and it will be dependent on the concentration of viral proteins and host receptors on the adhesive strips, which depends on the type of IRS discussed in Section III (b). These concentrations are estimated using the ligands, receptors, virus and viral detector dimensions found in \cite{martins2018computational}. 
 
\subsection{IRS Simulator}

We have performed simulations on Matlab to demonstrate the working of our proposed method. Our simulator comprises of a feedback-based beam steering mechanism. More specifically, we have used an Extremum Seeking Control (ESC) algorithm \cite{ashraf2020extremum} to steer the beam towards receiver. In this scheme, a dither signal (of 0.5 amplitude and 50 rad/s frequency in our case as shown in Table II) is used to excite the periodic response of the received signal power and corresponding change in the received power is used to drive and maintain the angle of reflection. For more details, readers are referred to \cite{ashraf2020extremum}. Moreover, the parameters for these simulations are listed in Table II.\\
Once the virus and its concentration have been detected by viral detection strips, this information is conveyed to the receiver by deviating the beam from its desired angle while at the same time beam steering mechanism tends to bring the beam back to its desired angle. 
\subsection{Simulation Results}

As indicated in Fig. \ref{fig:blockdia}, the information about virus detected via distributed sensors on IR-VD, is encoded into its reflection properties. This results  in  the  deviation  of  the beam  at  the  receiver and  gets notified about the presence of virus by decoding the pattern  given by the  changes occurring to  its  received signal power. Fig. \ref{fig:simsneeze} shows two simple simulated examples of the particle dispersion over a $1\times2$ m IR-VD placed on the wall with a varying distance of the sneezing person standing in front of the surface. In the figure, the heat map colour represents the density of droplets on the surface as the number of positive detections per tile, from low density (blue) to high (yellow).

When the virus information encoding process is integrated with our beam steering scheme \cite{ashraf2020extremum}, we enable communicating information about viral contamination.  In the figure we observe that a beam after reflecting through the IR-VD was initially targeting towards 48 degrees but the receiver was located at 50 degrees (this 50 degrees serves as a reference angle) with respect to the wall. Our beam steering algorithm works based on a feedback principle and tends to steer the beam towards 50 degrees at receiver location with  short transient flicking behavior of around 1 second. It can be observed that once the beam tracks the receiver at 50 degrees, at the time instant (1 s), the received signal power is at its maximum value of 1 due to normalization. At 2 sec time instant, the beam flicking process starts due to the viral detection. The flicking results in a small angle deviation based on the phase gradient change. This disturbance causes the angle to deviate by 1 degree and the beam reflects towards 49 degrees, and there is a corresponding decrease in the received power at the same instant. The pattern of change in the received signal  power and the timing between the beam flicks, conveys a message to the receiver. Indicatively, we have considered 5 bits of information (5 deviations)  
which are encoded at 2.5, 3.1, 4.1, 5.1 and 6.1 seconds. First two bits are used to represent the concentration and type of virus detected,  while the last three bits correspond to the density of virus that has bound onto the IR-VD. The time duration of the deviation corresponds to bit 0 or 1 as shown in the figures as well. For the one meter distance case, the width of the disturbance at 2.1, 3.1 and 6.1 s is double, corresponding to bit 1, compared to the time width of deviation at 4.1 and 5.1 second time instants, which corresponds to bit 0. 

\section{Challenges and Future Directions}

While the introduction of IRS will transform the management of wireless signals going forward into the future, the integration of viral detector and conveying that information to mobile devices will increase the power and need of this new infrastructure through the IR-VD. However, there are a number of challenges that will need to be overcome for the IR-VD to have its dual function. This section will review some of these challenges and future directions.

\subsection{Viral Detection Uncertainties}

Airborne virus detection is particularly challenging since it aims to obtain a sample of the viral particles to detect. Existing literature  \cite{fronczek2015biosensors} shows that these techniques need to consider a few fundamental system characteristics including particle random motion propagation, sample quantity variations on the reflector that can be due to low viral particle density in the aerosol particles, dust interference, and the affinity of virus binding to receptor structures on the adhesive strips. These challenges currently inhibit developments in airborne virus detector, which is reflected in the number of existing products available in the market today. At the same time, the environments will also require restricted disruptive air-flow to ensure that the virus can bind successfully onto the receptors and protect the sample for the duration of the detection. 
The usage of air trapping mechanisms can provide a protection to the air samples in the vicinity of the biosensing technology so that disruptive air-flow do not interfere with the binding process between virus and the sensing receptors. However, the design of the intelligent reflectors, whether they have different sizes and shapes of tiles, can possibly overcome these challenges, and further mathematical and testing work is most needed.

\subsection{Linking to Contact Tracing}
One way of countering the spread of airborne viruses is by social distancing. As introduced in Section \ref{sec:airborne_virus},  the recommended distance for people to stay apart from each other should be approximately two meters. Based on this, solutions that monitor the social distancing and identify the close contacts of known infected individuals are being proposed using the current wireless technologies, such as Bluetooth, although there are concerns in the accuracy of distance measurements between the devices. Our proposed viral detectors can complement such apps because our virus detection strips are integrated with a communication system that provides near real-time alerts to those in the vicinity of the virus particles. This can serve as another sensor information point for the contact tracing app that can propagate between the devices to avoid certain areas that may contain droplets.


\subsection{Virus Ultraviolet Treatment}
While the detectors can enable the virus to bind to provide information on hazardous regions, there also needs to be a solution to treat the virus to minimize physical contacts between people and the wall.  Recent research has found that using ultraviolet light can inactivate airborne viruses \cite{welch2018far}. A possible addition to the adhesive strips are local ultraviolet sources that are able to inactivate the virus once they have bound onto the receptors. The frequency range of the ultraviolet light are chosen in such way that it only affects the viral proteins and not the receptors on the strip. This ensures that the strips can be reused again to allow the receptors to become free, and at the same time ensure that the droplets do not infect any people within the vicinity. However, this can increase the complexity and cost of the IR-VD in order to ensure that the ultraviolet source will emit sufficient rays to treat the bound virus.


\subsection{Integration into the Environment}
A limitation of the proposed solution is that it will only detect the virus emitted through droplets in the vicinity of the reflectors. A next step would be the application of the viral detector strips to infrastructures within domestic and commercial environments, such as tables, floors and shopping store shelves. For that we need higher complexity beam control techniques that can overcome also signal losses from different object materials, such as the one proposed in \cite{pengnoo2020digital} for example. This will ensure very accurate detection of any virus droplets in areas of high density of people. However, a number of issues needs to be considered. This includes the task of transmitting signals from the strips to a device within the vicinity, which may require reflection of signals from a base station. Another challenge is the increase in cost of designing furniture, cabinets, and shelves that contain integrated circuits to connect to the adhesive strips containing the virus detectors as well as reflection communication system.  

\section{Conclusion}

{Novel viruses easily shared by air can bring the modern civilisation to an abrupt halt, as shown by the course of 2020. In indoor environments, viral load is often observed on surfaces, easily landing there through virus carriers breathing, sneezing, coughing or speaking. An indication of existence of such load in the room or its parts would be an important information for health and safety protocols in the building, an important health information for the people in the shared common space, and for general public health. Our design of IR-VD enables such pervasive sensing of viral presence, coupled with the roll-out of next generation wireless infrastructure, the IRS. The paper present a design that integrates such viral detectors through adhesive strips and how they can be integrated into conventional IRS. The communication of the viral load is achieved by manipulating the reflected signals through beam flicking that encodes the information about the virus bound on the call, and not compromising the orignal reflected signals. We have shown how such a system could operate, why is the convergence of technologies in this case natural and necessary, and what the future directions might look like for this synergy of communications and health technology very much in spirit of the sixth wireless generation.


%





\ifCLASSOPTIONcaptionsoff
  \newpage
\fi




\bibliography{bib/refe}

\begin{thebibliography}{10}
\providecommand{\url}[1]{#1}
\csname url@samestyle\endcsname
\providecommand{\newblock}{\relax}
\providecommand{\bibinfo}[2]{#2}
\providecommand{\BIBentrySTDinterwordspacing}{\spaceskip=0pt\relax}
\providecommand{\BIBentryALTinterwordstretchfactor}{4}
\providecommand{\BIBentryALTinterwordspacing}{\spaceskip=\fontdimen2\font plus
\BIBentryALTinterwordstretchfactor\fontdimen3\font minus
  \fontdimen4\font\relax}
\providecommand{\BIBforeignlanguage}[2]{{%
\expandafter\ifx\csname l@#1\endcsname\relax
\typeout{** WARNING: IEEEtran.bst: No hyphenation pattern has been}%
\typeout{** loaded for the language `#1'. Using the pattern for}%
\typeout{** the default language instead.}%
\else
\language=\csname l@#1\endcsname
\fi
#2}}
\providecommand{\BIBdecl}{\relax}
\BIBdecl

\bibitem{ziegler20206g}
V.~Ziegler and S.~Yrjola, ``{6G} indicators of value and performance,'' in
  \emph{2020 2nd {6G} Wireless Summit (6G SUMMIT)}.\hskip 1em plus 0.5em minus
  0.4em\relax IEEE, 2020, pp. 1--5.

\bibitem{Liaskos2018anewwireless}
C.~Liaskos, S.~Nie, A.~Tsioliaridou, A.~Pitsillides, S.~Ioannidis, and
  I.~Akyildiz, ``A new wireless communication paradigm through
  software-controlled metasurfaces,'' \emph{IEEE Communications Magazine},
  vol.~56, no.~9, pp. 162--169, 2018.

\bibitem{ashraf2020extremum}
N.~Ashraf, M.~Lestas, T.~Saeed, H.~Taghvaee, S.~Abadal, A.~Pitsillides, and
  C.~Liaskos, ``Extremum seeking control for beam steering using
  hypersurfaces,'' in \emph{2020 IEEE International Conference on
  Communications Workshops (ICC Workshops)}.\hskip 1em plus 0.5em minus
  0.4em\relax IEEE, 2020, pp. 1--6.

\bibitem{hurst2007manual}
S.~A. Sattar, N.~Bhardwaj, and M.~K. Ijaz, \emph{Airborne Viruses}.\hskip 1em
  plus 0.5em minus 0.4em\relax American Society for Microbiology Press, 2007,
  ch. 3.2.7, pp. 1--21.

\bibitem{fronczek2015biosensors}
C.~F. Fronczek and J.-Y. Yoon, ``Biosensors for monitoring airborne
  pathogens,'' \emph{Journal of laboratory automation}, vol.~20, no.~4, pp.
  390--410, 2015.

\bibitem{ijaz2016generic}
M.~K. Ijaz, B.~Zargar, K.~E. Wright, J.~R. Rubino, and S.~A. Sattar, ``Generic
  aspects of the airborne spread of human pathogens indoors and emerging air
  decontamination technologies,'' \emph{American Journal of Infection Control},
  vol.~44, no.~9, pp. S109--S120, 2016.

\bibitem{de2014role}
M.~de~Graaf and R.~A. Fouchier, ``Role of receptor binding specificity in
  influenza a virus transmission and pathogenesis,'' \emph{The EMBO journal},
  vol.~33, no.~8, pp. 823--841, 2014.

\bibitem{dai2020reconfigurable}
L.~Dai, B.~Wang, M.~Wang, X.~Yang, J.~Tan, S.~Bi, S.~Xu, F.~Yang, Z.~Chen,
  M.~Di~Renzo \emph{et~al.}, ``Reconfigurable intelligent surface-based
  wireless communications: Antenna design, prototyping, and experimental
  results,'' \emph{IEEE Access}, vol.~8, pp. 45\,913--45\,923, 2020.

\bibitem{li2019intelligent}
L.~Li, Y.~Shuang, Q.~Ma, H.~Li, H.~Zhao, M.~Wei, C.~Liu, C.~Hao, C.-W. Qiu, and
  T.~J. Cui, ``Intelligent metasurface imager and recognizer,'' \emph{Light:
  Science \& Applications}, vol.~8, no.~1, pp. 1--9, 2019.

\bibitem{naqvi2019beam}
A.~H. Naqvi and S.~Lim, ``A beam-steering antenna with a fluidically
  programmable metasurface,'' \emph{IEEE Transactions on Antennas and
  Propagation}, vol.~67, no.~6, pp. 3704--3711, 2019.

\bibitem{khalid2019communication}
M.~Khalid, O.~Amin, S.~Ahmed, B.~Shihada, and M.-S. Alouini, ``Communication
  through breath: Aerosol transmission,'' \emph{IEEE Communications Magazine},
  vol.~57, no.~2, pp. 33--39, 2019.

\bibitem{huang2019rising}
Y.~Huang, X.~Chen, M.~Wen, L.-L. Yang, C.-B. Chae, and F.~Ji, ``A rising
  edge-based detection algorithm for mimo molecular communication,'' \emph{IEEE
  Wireless Communications Letters}, vol.~9, no.~4, pp. 523--527, 2019.

\bibitem{martins2018computational}
D.~P. Martins, M.~T. Barros, M.~Pierobon, M.~Kandhavelu, S.~Balasubramaniam
  \emph{et~al.}, ``Computational models for trapping ebola virus using
  engineered bacteria,'' \emph{IEEE/ACM Transactions on Computational Biology
  and Bioinformatics}, vol.~15, no.~6, pp. 2017--2027, 2018.

\bibitem{welch2018far}
D.~Welch, M.~Buonanno, V.~Grilj, I.~Shuryak, C.~Crickmore, A.~W. Bigelow,
  G.~Randers-Pehrson, G.~W. Johnson, and D.~J. Brenner, ``{Far-UVC} light: A
  new tool to control the spread of airborne-mediated microbial diseases,''
  \emph{Scientific Reports}, vol.~8, no.~1, pp. 1--7, 2018.

\bibitem{pengnoo2020digital}
M.~Pengnoo, M.~T. Barros, L.~Wuttisittikulkij, B.~Butler, A.~Davy, and
  S.~Balasubramaniam, ``Digital twin for metasurface reflector management in 6g
  terahertz communications,'' \emph{IEEE Access}, vol.~8, pp.
  114\,580--114\,596, 2020.

\end{thebibliography}
\bibliographystyle{IEEEtran}
\end{document}